# Non-Equilibrium Phase Transitions and Domain Walls

D.Coulson[1], Z.Lalak[1,2], B.Ovrut[1]
[1]Department of Physics
University of Pennsylvania
Philadelphia, PA 19104
[2]Institute of Theoretical Physics
University of Warsaw
Warsaw, Poland 00-681

**Abstract**

Non-equilibrium phase transitions of a scalar field in an expanding spacetime are discussed. These transitions are shown to lead, for appropriate potential energy functions, to a biased choice of vacuum structure which can be analytically described using percolation theory. The initial domain wall networks, which form between different vacua, are evolved in time by computer simulation and their behavior in time analyzed. It is shown that, unlike systems in thermal equilibrium, domain walls formed in biased systems persist for only a short time before decaying exponentially away. This result opens the door to a complete re-analysis of domain walls in cosmology.

## 1 The "Structuron" Field

We begin by considering a real scalar field, $\phi$, which lives in an expanding Friedman-Robinson-Walker (FRW) spacetime. We will assume that this field interacts with other fields by gravitation alone, and interacts with itself via a potential energy of the form

$$V(\phi) = V_0 \left( \cos \frac{\phi}{v} + 1 \right), \tag{1}$$

where

$$\frac{V_0}{v^2} \ll H_i^2, \tag{2}$$

$$V_0 \ll v^4,$$

and $H_i$ is the initial Hubble parameter during the epoch of inflation. One minimum of this potential occurs at $\phi_0 = \pi v$. Expanding $V$ around this minimum gives

$$V = \frac{1}{2} \left( \frac{V_0}{v^2} \right) \phi^2 + \frac{1}{4!} \left( \frac{V_0}{v^4} \right) \phi^4 + ... \tag{3}$$

Defining mass $m \equiv \frac{V_0}{v^2}$ and coupling $\lambda \equiv \frac{V_0}{v^4}$, then the conditions in equation (2) imply that

$$m \ll H_i, \tag{4}$$

$$\lambda \ll 1.$$

It follows that field $\phi$ is not in thermal equilibrium. Many such fields appear naturally in various "beyond the standard model" scenarios. Here we will simply discuss such a field in the abstract, referring to it by the generic name "structuron" since, as we will see, it can play a role in creating large scale extra-galactic structure. The behavior of a structuron is determined by considering the quantum mechanics of a real scalar quantum field, $\hat{\phi}$, in expanding FRW spacetime. This quantum field can be expanded as

$$\hat{\phi} = \phi_c \hat{1} + \hat{\phi}_q. \tag{5}$$

Field $\phi_c$ satisfies the classical equation of motion

$$\ddot{\phi}_c + 3H\dot{\phi}_c + \frac{\partial V}{\partial \phi_c} = 0. \tag{6}$$

During inflation, when $H = H_i$, and long afterward during the expansion epoch, the conditions in equations (4) imply that one can drop the potential energy term to leading order. The result is that, to leading order

$$\phi_c = \mathcal{V} \tag{7}$$

where $\mathcal{V}$ is a completely arbitrary constant. To the next order, the classical field develops a tiny damped velocity

$$\dot{\phi}_c \cong \frac{V}{Hv} \tag{8}$$

which can safely be ignored. The quantum operator, $\hat{\phi}_q$, and its physical effect, is a more complicated issue. It is well known that the quantum fluctuations associated this field lead, during the inflationary epoch, to the formation of a "weakly inhomogeneous, quasi-classical" scalar field, $\Phi$. This consists, just at the end of inflation, of a quantum induced zero mode, plus a part which can be Fourier decomposed into wavelengths satisfying the condition

$$H_i^{-1} \leq \lambda \leq L. \tag{9}$$

Here $H_i^{-1}$ is the radius of the event horizon at the end of inflation and L is the radius of the Universe. During the expansion phase, the radius of the event horizon increases with time. At redshift $z$, this radius is given by $l_c(z) = H(z)^{-1}$. As wavelengths of the quasi-classical field come within the growing horizon, they rapidly decay in amplitude and can be ignored. At any given redshift $z$, only those wavelengths satisfying

$$l_c(z) \leq \lambda \leq L \tag{10}$$

remain "frozen" and compose the quasi-classical field. Field $\Phi$, at any redshift $z$, has an interesting spatial decomposition that will be useful in our later analysis. To discuss this we first must introduce the two-point correlation function, $\xi(l)$, defined by

$$\xi(l) \equiv \langle 0 | \hat{\phi}_q(x+l) \hat{\phi}_q(x) | 0 \rangle \tag{11}$$

where the wavelengths in the quantum operators are to be cut off as in equation (10) and $l$ is arbitrary. The translational invariance of the Bunch-Davies vacuum $|0\rangle$ implies that $\xi$ is independent of coordinate $x$. Expression (11) can be evaluated, and for $l_c \ll l \ll L$, we find that

$$\xi(l) \cong \frac{H_i^2}{4\pi^2} \ln \frac{L}{l}. \tag{12}$$

This result is completely dominated by the long wavelength modes satisfying $l \leq \lambda \leq L$. Secondly, we must introduce a measure of the random fluctuations between two points, $\Delta(l)$, which is defined by

$$\Delta(l) = \frac{H_i^2}{4\pi^2} \ln \frac{l}{l_c}. \tag{13}$$

This function is completely dominated by the short wavelength modes satisfying $l_c(z) \leq \lambda \leq l$. Note that

$$\xi(l) + \Delta(l) = \frac{H_i}{4\pi^2} \ln \frac{L}{l_c} \tag{14}$$

which is the mean square value of all the fluctuations in $\Phi$. We are now in a position to decompose field $\Phi$, as alluded to above. Consider the universe at some fixed redshift $z$. Let us "coarse-grain" the universe into spheres of radius $l$. To first approximation, all of these spheres are correlated by the longer wavelength modes, which they share in common. Therefore, they all share an approximate zero mode background, $\mathcal{B}_l$, defined by

$$\mathcal{B}_l = +\sqrt{\xi(l)}. \tag{15}$$

However, within each sphere there are fluctuations due to the shorter wavelength modes. Since these fluctuations arise in a causally independent way from one sphere to the next, each sphere exhibits an independent fluctuation, $\mathcal{F}_l$, whose value is determined probabilistically from the Gaussian distribution function

$$P(\mathcal{F}_l) = \frac{1}{\sqrt{2\pi\Delta(l)}} e^{-\frac{\mathcal{F}_l^2}{2\Delta(l)}}. \tag{16}$$

Note that the root mean square value of this distribution is given by

$$\text{rms}(\mathcal{F}_l) = +\sqrt{\Delta(l)}. \tag{17}$$

Unlike $\mathcal{B}_l$ which is a spatial constant, $\mathcal{F}_l$ varies randomly over the coarse-grained manifold. It follows that for any length scale $l$, the weakly inhomogeneous, quasi-classical scalar field can be decomposed as

$$\Phi(x) = \mathcal{B}_l + \mathcal{F}_l(x). \tag{18}$$

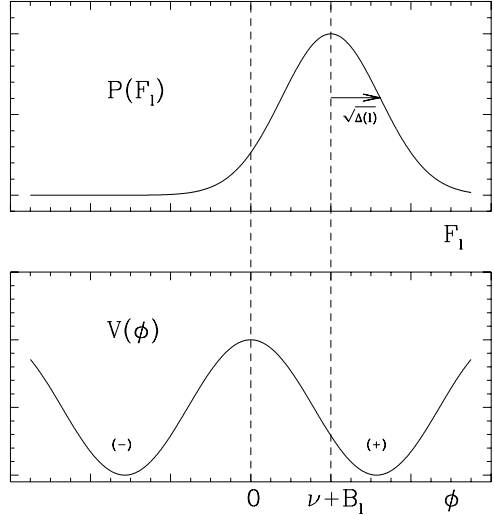

Figure 1: Distribution function, $P(\mathcal{F}_l)$ and potential, $V(\phi)$.

At any point in space, the relationship between the total shared zero mode background $\mathcal{V} + \mathcal{B}_l$, the quantum fluctuation $\mathcal{F}_l$ and the potential energy is shown in Figure 1. As long as the potential energy in equation (6) can be ignored, no phase transition takes place. However, after a long time, the redshift is such that the kinetic energy density becomes comparable, and then smaller than, the potential energy. Let us say that they become comparable, and the potential energy no longer ignorable, at redshift $z_t$. At this time, at each point in space, the field must roll either toward vacuum $(+)$ or toward vacuum $(-)$. It is clear from Figure 1 that the the probability that it roles to the $(+)$ vacuum is given by

$$p = \int_0^\infty d\mathcal{F}_l P(\mathcal{F}_l). \qquad (19)$$

It is important to note that, since classical background $\mathcal{V}$ is arbitrary, $p$ can take any value in the range $0 \leq p \leq 1$. In general, $p \neq 1/2$. Similarly, the probability that the field rolls to the $(-)$ vacuum is given by $1 - p$. Putting everything together, we see that at the time of the phase transition, $z_t$, in any spatial sphere of radius $l$ (that is, at any coarse-grained point) the system settles into vacuum $(+)$ with arbitrary probability $p$, or vacuum ($-$) with probability $1 - p$, and that the choice of the vacuum in any two spheres is statistically independent [1]. Now between any two neighboring spheres which have a different vacuum choice, the scalar field $\phi$ will form a topological kink; that is, a domain wall. Hence, at $z_t$ an initial domain wall network will form over all space. Can we determine the spatial structure of this initial domain wall network? Yes, using percolation theory.

## 2 Percolation Theory

Percolation theory predicts statistical characteristics and topological properties of a "typical" vacuum pattern formed on a lattice. If we partition space into a cubic lattice, with each cube having volume $l^3$, then percolation theory becomes applicable to the prediction of the initial distribution of domain walls. In the following, we will always restrict our discussion to three dimensional, cubic lattices. Percolation theory tells us that there exists some critical probability, $p_c$. If the population probability for a vacuum exceeds this critical value, that vacuum will percolate the lattice; that is, one can trace a path from one face of the lattice to another without crossing a domain wall. If the bias probability of a vacuum is less than the critical value, however, that vacuum will not percolate, and domain wall "bags", or clusters, of this vacuum will form. Whether one, both or neither vacua percolate depends on the relative values of $p$, $1 - p$ and $p_c$.

It has been shown that $p_c = 0.311$ is the critical probability for a cubic lattice in three dimensions [2]. Thus, for $p < p_c$, the (+) vacua are in finite clusters while the ($-$) vacua sites lie predominantly in a large percolating cluster, since necessarily $1 - p > p_c$. It follows that the associated domain walls are relatively small, forming around the compact boundaries of the finite clusters. Figure 2 shows this behavior in the case of a small bias probability, $p = 0.1 < p_c$. Note that there exist only finite, disconnected domain wall bags, the number and size of which can be shown to grow as $p$ approaches $p_c$ from below. For the case that both $p$ and $1 - p$ exceed $p_c$, both vacua will percolate the lattice, and infinite (that is, lattice sized) domain walls separating the vacua will form. A small number of domain wall bags, disconnected from the percolating cluster, also form but their number decreases as $p$ approaches 0.5. Figure 3 shows this "infinite" domain wall structure in the limiting case of $p = 0.5$. The crucial lesson to be learned from all this is that when both vacua percolate, the topology of the post-transition vacuum is one of long, convoluted domain walls stretching across all space, whereas, when $p < p_c$, the vacuum is composed of small, compact domain wall bags.

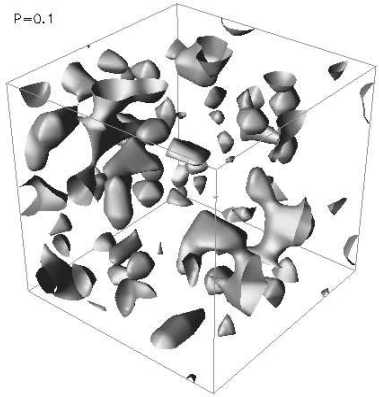

Figure 2: Initial distribution of domain walls in three dimensions with bias probability, $p = 0.1$. Here we show a $(10\Lambda)^3$ grid.

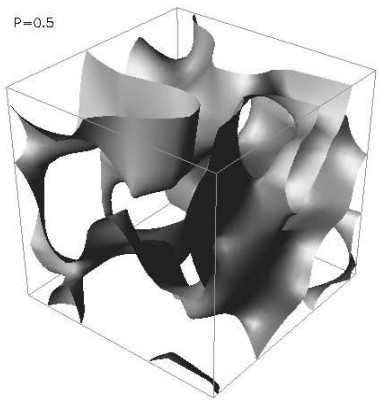

Figure 3: Initial distribution of domain walls in three dimensions with bias probability, $p = 0.5$. For clarity we show here only a $(5\Lambda)^3$ grid.

Percolation theory allows one to give a reasonably accurate mathematical description of the number of finite $s$-clusters, their radius and the size of their boundary. Here $s$ denotes the number of neighboring lattice sites that are occupied by the same vacuum. Let $n_s$ be the probability that a given lattice site is an element of an $s$-cluster. This is a fundamental quantity, given by the ratio of the total number of s-clusters, $N_s$, over the total number of lattice sites, $N$. An analytical expression for this quantity has been found using scaling arguments and Monte Carlo simulations [3]. The result is

$$n_s = .0501 s^{-\tau} \, e^{-.6299(\frac{p-p_c}{p_c})s^\sigma [(\frac{p-p_c}{p})s^\sigma + 1.6679]}, \tag{20}$$

where $\tau = 2.17$ and $\sigma = .48$. Similarly, the average radius of gyration for an s-cluster $R_s(p)$, for $p < p_c$ and $s > s_\xi$, is found to be

$$R_s(p) = .702(p_c - p)^{.322} s^{.55} \Lambda, \tag{21}$$

where

$$s_\xi = \left(\frac{.311}{|p - .311|}\right)^{2.08}. \tag{22}$$

It can also be shown, for $p < p_c$ and $s \gtrsim 5$, that every $s$-cluster has a boundary composed of

$$t_s = \left(\frac{1-p}{p}\right) s \tag{23}$$

sites. One can easily check using the formula for $n_s$ that, on a given lattice, the number of $s$-clusters falls rather quickly with growing $s$. Hence there is an $s_{max}$ such that the total number of $s_{max}$-clusters is 1. In other words, formation of clusters with $s$ much larger than $s_{max}$ is extremely improbable. This means that on a given lattice there exists an upper statistical cut-off on the size of observable clusters..

Using the above formulae, as well as other results found in [3], it is possible, for a fixed lattice and a given value of $p$, to compute the surface area for the domain walls between the $(+)$ and the $(-)$ vacua. In particular, we have attempted to compute the domain wall surface area for the three dimensional cubic lattice case. However, this calculation is very difficult when there are two percolating vacua and hence, we are restricted to a calculation for $p < p_c$. This difficulty arises from the fact that the associated large domain walls are highly fractalized and hence their surface area is difficult to characterize. The finite size of the lattice further reduces the value of $p$ for which we can perform a reasonable calculation to $p \lesssim 0.25$. Furthermore, formula (20) for $n_s$ is not very accurate for $p < 0.175$. This puts a lower bound on the calculation. A very reasonable calculation can be performed in the range $0.175 \leq p \leq 0.25$ but, even in this range, we estimate an error of about 10%. Our results for the domain wall surface area are plotted, along with the 10% error bars, as curve (a) in Figure 4. These results can be checked, and extended to any values of $p$, simply by letting the computer evaluate wall surface area. The solid line (b) in Figure 4 represents the mean initial surface area generated for $0 \leq p \leq 0.5$. The initial area grows monotonically with $p$, approaching the maximal value of $1.5 \times V$ at $p = 0.5$. Note that in the region where we can compare the percolation prediction with the computer experiment, there is good agreement.

We conclude that, by using percolation theory, we can analytically determine the structure of the initial network of domain walls. The principal result is that for $p < p_c$ the domain walls form small, compact bags whereas when $p \geq p_c$ the domain walls become very long and stretch across the entire universe. These results have been verified, and made numerically more accurate, using computer simulations. However, the initial domain wall networks do not satisfy the static equation of motion and, therefore, must evolve in time. The role of dynamics in propagating or erasing these initial conditions is discussed in the following section.

## 3  Evolution of Domain Walls

To investigate the evolution of the initial domain wall networks described above, we choose to follow Press, Ryden and Spergel (hereafter PRS)[4]. The dynamics of the scalar field, $\phi$, are determined by the equation of motion

$$\frac{\partial^2 \phi}{\partial \eta^2} + \frac{2}{\eta}\frac{d \ln a}{d \ln \eta}\frac{\partial \phi}{\partial \eta} - \nabla^2 \phi = -a^2 \frac{\partial V}{\partial \phi}, \qquad (24)$$

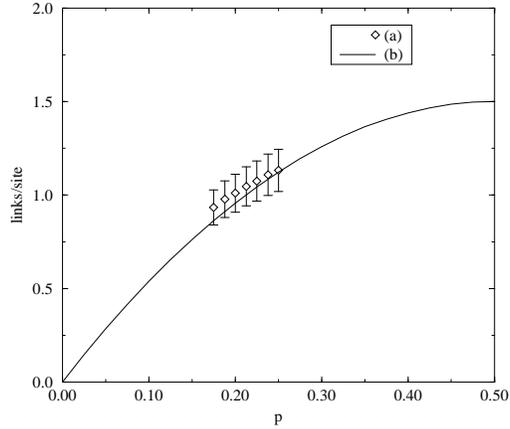

Figure 4: The number of links across which a domain wall falls per lattice site. Points (a) are the $3d$ percolative predictions in the regime $0.175 \leq p \leq 0.25$, with their associated uncertainty ($\sim 10\%$). Line (b) shows the measured values (averaged over 20 three dimensional samples) for $0 \leq p \leq 0.5$.

where $\eta$ is the conformal time co-ordinate, $a$ is the scale factor of the universe ($a \sim \eta$ in the radiation era, and $a \sim \eta^2$ in the matter era), $V$ is the scalar potential and the spatial gradients are with respect to co-moving co-ordinates. The scalar potential determines the topology of the vacuum manifold. Instead of the periodic potential given in 1, we will use the quartic potential

$$V(\phi) = V_0 \left( \frac{\phi^2}{\phi_0^2} - 1 \right)^2. \qquad (25)$$

This potential has two degenerate vacua, $\phi = \pm \phi_0$, separated by a potential barrier $V_0$ and closely approximates the cosine potential in the region between any two minima.

Following PRS we can define a physical domain wall thickness $w_0$ given by

$$w_0 \equiv \pi \frac{\phi_0}{\sqrt{2V_0}}. \qquad (26)$$

The ratio of the wall thickness to the horizon size at the time of the phase

transition
$$W_0 \equiv \frac{w_0}{a(\eta_0)}\frac{1}{\eta_0}\frac{d\ln a}{d\ln \eta}\bigg|_{\eta_0} \qquad (27)$$
then sets $\eta_0$, the conformal time of the phase transition and the time at which we begin the simulation.

In order to avoid some technical difficulties, we will use a generalization of equation (24) given by
$$\frac{\partial^2 \phi}{\partial \eta^2} + \frac{\alpha}{\eta}\frac{d\ln a}{d\ln \eta}\frac{\partial \phi}{\partial \eta} - \nabla^2 \phi = -a^\beta \frac{\partial V}{\partial \phi}. \qquad (28)$$

Henceforth, $\alpha = \beta = 2$ which reproduces equation (24) will be replaced with $\alpha = 3$, $\beta = 0$. We refer the reader to PRS for a full justification of this change.

We evolve equation (28) on a regular Cartesian grid with periodic boundary conditions. Our finite difference scheme is second order accurate in both space and time, with the lattice equations

$$\delta \equiv \frac{1}{2}\alpha \frac{\Delta\eta}{\eta}\frac{d\ln a}{d\ln\eta}, \qquad (29)$$

$$\left(\nabla^2 \phi\right)_{i,j,k} \equiv \phi_{i+1,j,k} + \phi_{i-1,k,k} + \phi_{i,j+1,k} + \phi_{i,j-1,k}$$
$$+ \phi_{i,j,k+1} + \phi_{i,j,k-1} - 6\phi_{i,j,k}, \qquad (30)$$

$$\dot{\phi}_{i,j,k}^{n+1/2} = \frac{(1-\delta)\dot{\phi}_{i,j,k}^{n-1/2} + \Delta\eta\left(\nabla^2 \phi_{ijk}^n - a^\beta \frac{\partial V}{\partial \phi_{ijk}^n}\right)}{1+\delta} \qquad (31)$$

$$\phi_{i,j,k}^{n+1} = \phi_{i,j,k}^n + \Delta\eta \dot{\phi}_{i,j,k}^{n+1/2}. \qquad (32)$$

Here, subscripts refer to the spatial lattice co-ordinate, superscripts refer to the time co-ordinate, and $\dot{\phi} \equiv \partial\phi/\partial\eta$. The spatial grid size will be chosen to be $\Delta x = 1$. In this paper, we will set $\phi_0 = 1$. The scalar field initial conditions are then chosen using the prescription described above for various bias probabilities, $p$. That is, in the following we will use percolation theory with allowed field values of $\pm 1$ at any lattice site. We will also choose the initial field "velocity", $\dot{\phi}$, to be zero everywhere on the lattice.

Simulations were run in the radiation dominated epoch, with $a = (\eta/\eta_0)$ and the initial time, $\eta_0 = 1$. We chose a wall thickness $w_0 = 5$, making the ratio $W_0 = 5$. This value was used to ensure that the wall thickness was well above the lattice resolution scale (recall $\Delta x = 1$), while ensuring that for most of the dynamic range of the simulation, the wall–wall separation exceeded the wall thickness. The results are the following.

We present the evolution of the energy density of the network of domain walls in the radiation dominated epoch. As each simulation is evolved, the comoving wall area is determined (according to the prescription of PRS) and a plot of this area, $A$, per co-moving volume $V$ versus conformal time is produced. The simulations are run on a cubic lattice with $L^3$ sites where $L = 128$. They were run until a time $\eta = 128$, or until no more walls remained in the box. The domain wall thickness was again set to $w_0 = 5$, leaving us with only modest dynamical range in which to follow the network evolution.

A plot of $A/V$ for these runs is shown in Figure 5. The self-similar evolution seen in PRS for the $p = 0.5$ case is well reproduced in the time range $2w_0 < \eta < L/2$. Fitting the scaling portion ($10 < \eta < 64$) of the curve to the power law

$$A/V \propto \eta^{\bar{\nu}}, \tag{33}$$

we find $\bar{\nu} = -0.89 \pm 0.06$. This is to be compared with the value found in PRS of $\bar{\nu} = -0.92 \pm 0.06$ for their three dimensional simulations in a matter-dominated epoch.

Moving away from the $p = 0.5$ case, one sees a dramatic departure from self-similar scaling. In each case there is an exponential cut-off in the ratio $A/V$ at some characteristic time. For the cases of $p$ close to $1/2$, that is for $0.49 \leq p < 0.5$, we find that the curves are well fitted by a function of the form

$$A/V \propto \eta^{\bar{\nu}} e^{-\eta/\bar{\eta}}. \tag{34}$$

However, for the cases where $p < 0.49$ a simple exponential suffices

$$A/V \propto e^{-\eta/\bar{\eta}}. \tag{35}$$

Values for $\bar{\nu}$ and $\bar{\eta}$, averaged over 5 runs for each value of $p$, are given in Table 1.

For $p$ close to 0.5 the domain wall network appears to enter a quasi-scaling regime in which $A/V$ scales $\propto \eta^{\bar{\nu}}$, before eventually being exponentially cut-off at $\eta = \bar{\eta}$. In particular, for the bias $p = 0.499$ we see that the network scales exactly as the $p = 0.5$ case in the epoch $2w_0 \leq \eta \leq 30$, before the exponential decay is established. As $p \to p_c = 0.311$, however, $\bar{\eta}$ rapidly approaches the resolution size of the grid, and no evidence of early scaling is seen. This behavior continues as $p$ drops below the critical threshold.

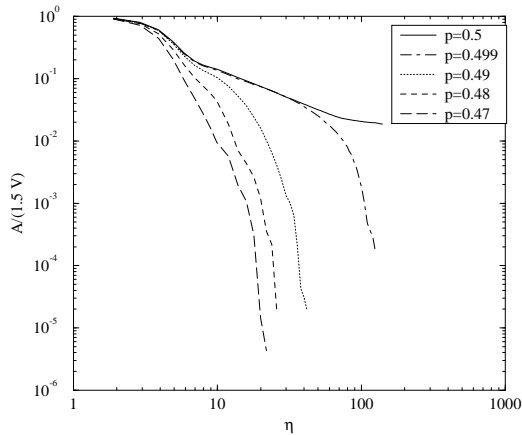

Figure 5: Evolution of the comoving area, $A$, of domain walls per volume, $V$, with conformal time, $\eta$, in three dimensional runs. For each bias, $p$, we show the evolution of one realization.

| $p$ | $\bar{\nu}$ | $\bar{\eta}$ |
|---|---|---|
| 0.5 | -0.89 | – |
| 0.499 | -0.43 | 38.7[†] |
| 0.49 | – | 5.4 |
| 0.48 | – | 2.8 |
| 0.47 | – | 1.8 |

Table 1: Fits to the plots of $A/V$ against $\eta$ for different initial bias probabilities, $p$, in three dimensions using the functional forms (34) and (35) given in the text. [†]: for the $p = 0.499$ case we report the best fit to equation (34), although the late time decay was found to be somewhat steeper than an exponential.

To conclude, in the three-dimensional simulations we see persistent scaling behavior precisely at $p = 0.5$. For $p$ below 0.5 but above $p \simeq 0.49$, we see scaling for a finite time which is then exponentially cut-off at some conformal time $\bar{\eta}$. The value of $\bar{\eta}$, which becomes very large as $p \to 0.5$, decreases rapidly as $p$ approaches 0.49. For $p$ below this value no scaling behavior is seen and the behavior is well described by a simple exponential for all conformal time.

## 4 Discussion

We conclude that domain walls formed during a biased non-equilibrium phase transition do not, in general, persist in time. For $p \neq 1/2$ they scale for a relatively short time before decaying away exponentially. Only when $p = 1/2$ exactly does persistent scaling behavior set in. It follows that biased domain walls evade the "no-go" arguments that apply to walls in thermal equilibrium. Since domain walls exhibit localized energy density, they will act as the seeds for the formation of large scale structure. A first attempt to analytically calculate the precise form of this structure, using percolation theory and the spherical matter collapse model, has been presented in [5]. The lesson learned there is that domain wall induced structure leaves a distinctive signature which could well open a cosmological window onto the microscopic structure of particle physics.

## Acknowledgments

We thank Paul Steinhardt, Rob Crittenden and David Spergel for their useful comments. The work of D.Coulson was supported by the DOE under Contract No. DOE-EY-76-C-02-3071, while that of Z.Lalak and B.Ovrut was supported in part by the DOE under Contract No. DOE-AC02-76-ERO-3071 and NATO Contract Ref. CRG 940784.